\documentclass[journal,onecolumn]{IEEEtran}
\IEEEoverridecommandlockouts
\usepackage{cite}
\usepackage[font=footnotesize,caption=false,subrefformat=parens,labelformat=parens]{subfig}
\usepackage{amsmath,amssymb,amsfonts}
\usepackage{algorithmic}
\usepackage{graphicx}
\usepackage{textcomp}
\usepackage{xcolor}
\usepackage{blindtext}
\def\BibTeX{{\rm B\kern-.05em{\sc i\kern-.025em b}\kern-.08em
    T\kern-.1667em\lower.7ex\hbox{E}\kern-.125emX}}

\usepackage{bbm}    
\begin{document}

\title{Outdoor mmWave Base Station Placement: A Multi-Armed Bandit Learning Approach}

\author{%
\normalsize Fatih Erden, Chethan K. Anjinappa, Ender Ozturk, and Ismail Guvenc%
\thanks{All the authors are with the Department of Electrical and Computer Engineering, North Carolina State University, Raleigh, NC 27606 (e-mail:\{ferden, canjina, eozturk2, iguvenc\}@ncsu.edu).}
}

\maketitle

\begin{abstract}
Base station~(BS) placement in mobile networks is critical to the efficient use of resources in any communication system and one of the main factors that determines the quality of communication. Although there is ample literature on the optimum placement of BSs for sub-6~GHz bands, channel propagation characteristics, such as penetration loss, are notably different in millimeter-wave~(mmWave) bands than in sub-6~GHz bands. Therefore, designated solutions are needed for mmWave systems to have reliable quality of service~(QoS) assessment. This article proposes a multi-armed bandit~(MAB) learning approach for the mmWave BS placement problem. The proposed solution performs viewshed analysis to identify the areas that are visible to a given BS location by considering the 3D geometry of the outdoor environments. Coverage probability, which is used as the QoS metric, is calculated using the appropriate path loss model depending on the viewshed analysis and a probabilistic blockage model and then fed to the MAB learning mechanism. The optimum BS location is then determined based on the expected reward that the candidate locations attain at the end of the training process. Unlike the optimization-based techniques, this method can capture the time-varying behavior of the channel and find the optimal BS locations that maximize long-term performance. \looseness=-1
\end{abstract}


\section*{Introduction}
\label{sec:Intro}
With the introduction of 5G mobile networks, many new features will be staged by the service providers as this new technology offers the highest wireless data rates ever. Exponential growth in data demand can only be met by sufficient amount of spectrum allocation, such that channel bandwidth on the order of gigahertz is needed to provide the envisaged speeds~\cite{3GPP}. Sub-6~GHz bands have high coverage probability in both indoor and outdoor environments; however, extreme spectrum shortage in these traditional bands reduces the appeal. Mostly-vacant millimeter-wave~(mmWave) frequencies come up at this point as a potential solution. For example, mmWave spectrum at 28~GHz and 39~GHz bands have already been licensed for mobile usage in the US. These bands, however, come with challenges as well as opportunities. The major challenge is the high loss rates in terms of both free-space path loss and penetration loss~\cite{khawaja2019coverage,erden201928}. Smart deployment of base stations~(BSs) is a possible solution to alleviate these drawbacks.\looseness=-1

Careful planning of the BS locations can help reduce the infrastructure costs while keeping the quality of service~(QoS) of communication at a desired level. BS placement problem is particularly challenging at mmWave frequencies and critical to solve for the commercial success of 5G deployments. Optimum placement of the BSs has been studied extensively for sub-6~GHz cellular frequencies~\cite{wong2006base,aldajani2008,lee2015base}. Since the function to be optimized is often non-convex, only a limited number of optimization techniques --subject to various assumptions, such as considering 2D environments-- have been attempted~\cite{Wright_1998}. However, channel propagation characteristics of mmWave bands are significantly different from those of the sub-6~GHz frequencies, making the available solutions for the latter impracticable for the former. 

\begin{figure}[t]
\centerline{\includegraphics[width=0.65\columnwidth]{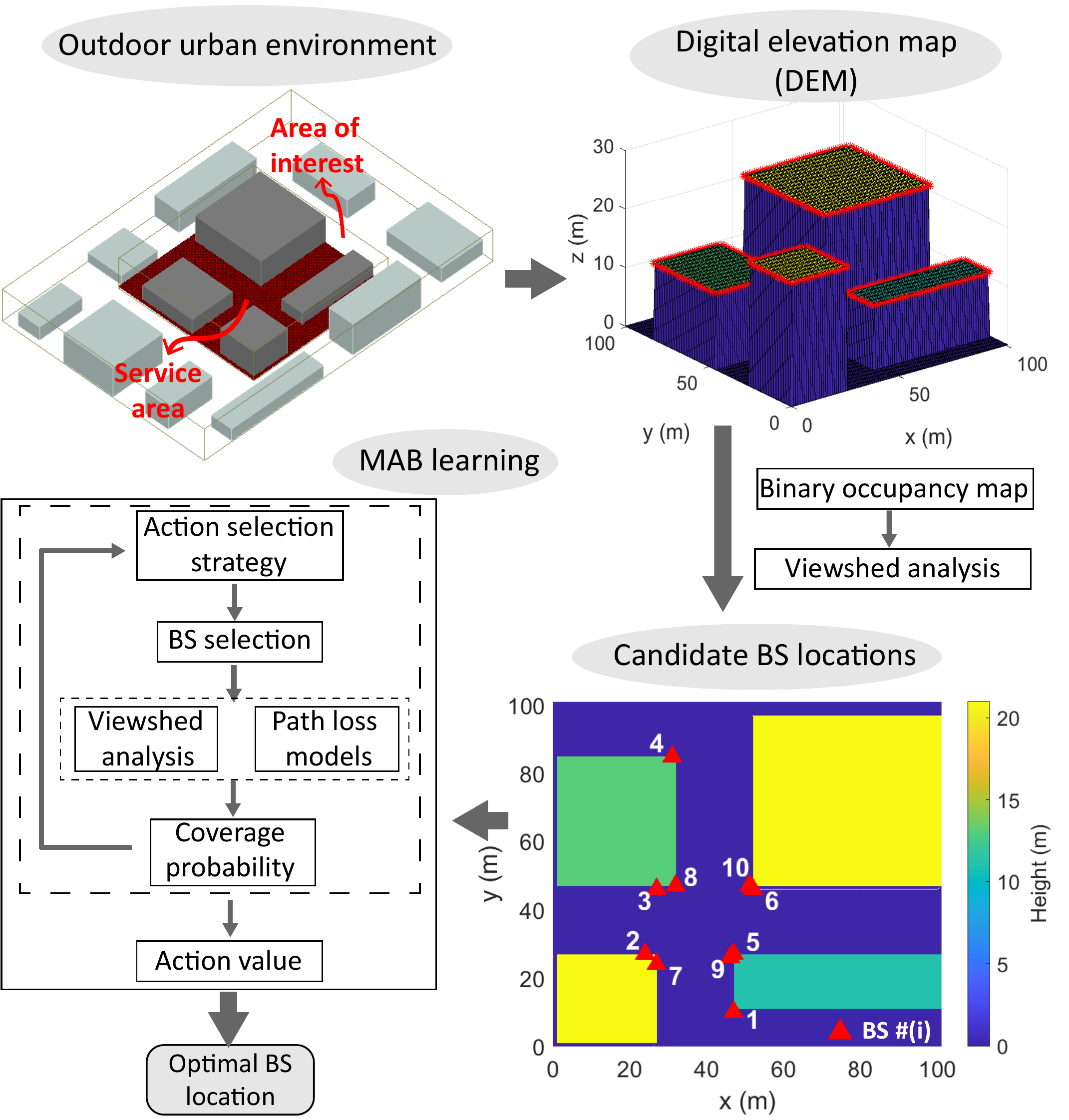}}
\caption{The flow-diagram of the MAB learning-based mmWave BS placement method.}
\label{fig:process}
\end{figure}

There are a few studies aiming to solve the BS deployment problem for mmWave bands. Most of these studies assume the environment to be in 2D~\cite{Xu_only2D,yanik2016, ioannis_ITS_2019,Yanikomeroglu_2017}. However, as it will be discussed in the next section, mmWave communication is easy to suffer from the blockages. In addition, the typical range of a mmWave BS at frequencies 28$-$73~GHz is between 100$-$200~m~\cite{Rappa_200m_range}. Therefore, especially in urban areas, the environment and the optimum BS locations should be defined in 3D. Studies which assume 3D environments usually take the problem as a pure optimization problem and/or offer a solution for an over-simplified environment~\cite{Cao_2017_pureopt,Abdel_Rahman_3D_but_nomap,Wang_3D_but_nomap}. Besides, in the optimization-based techniques, mean statistics of the channel are used for the evaluation of a given performance metric, and hence, the time-varying behavior of the channel (due to, e.g., mobile blockages and nonuniform user distribution) are often overlooked.

This article proposes a multi-armed bandit~(MAB) learning approach to determine the optimal locations of mmWave BSs in urban environments.
Given such an environment with a predefined service area and possible BS locations, a digital elevation model~(DEM) is generated first, followed by the extraction of binary occupancy map~(BOM). Candidate BS locations are determined by performing viewshed analysis for each possible BS location. Coverage performance associated with the candidate locations is assessed by the MAB learning through time and the BS location that is optimal in the sense of maximizing the long-term coverage probability is identified. The overall process is summarized in Fig.~\ref{fig:process}. Two key contributions of this article can be highlighted as follows. Firstly, it estimates the coverage probability using a model that incorporates viewshed analysis and path loss models for mmWave channels. Secondly, it takes the temporal changes in the channel (e.g., blockage probability) into account and finds the BS location that is the most beneficial to the long-term communication performance.\looseness = -1

\begin{figure}[t]
\vspace{-3mm}
\centering
\subfloat[]{\includegraphics[trim=0.2cm 0cm 0cm 0.6cm, clip,width=0.51\linewidth]{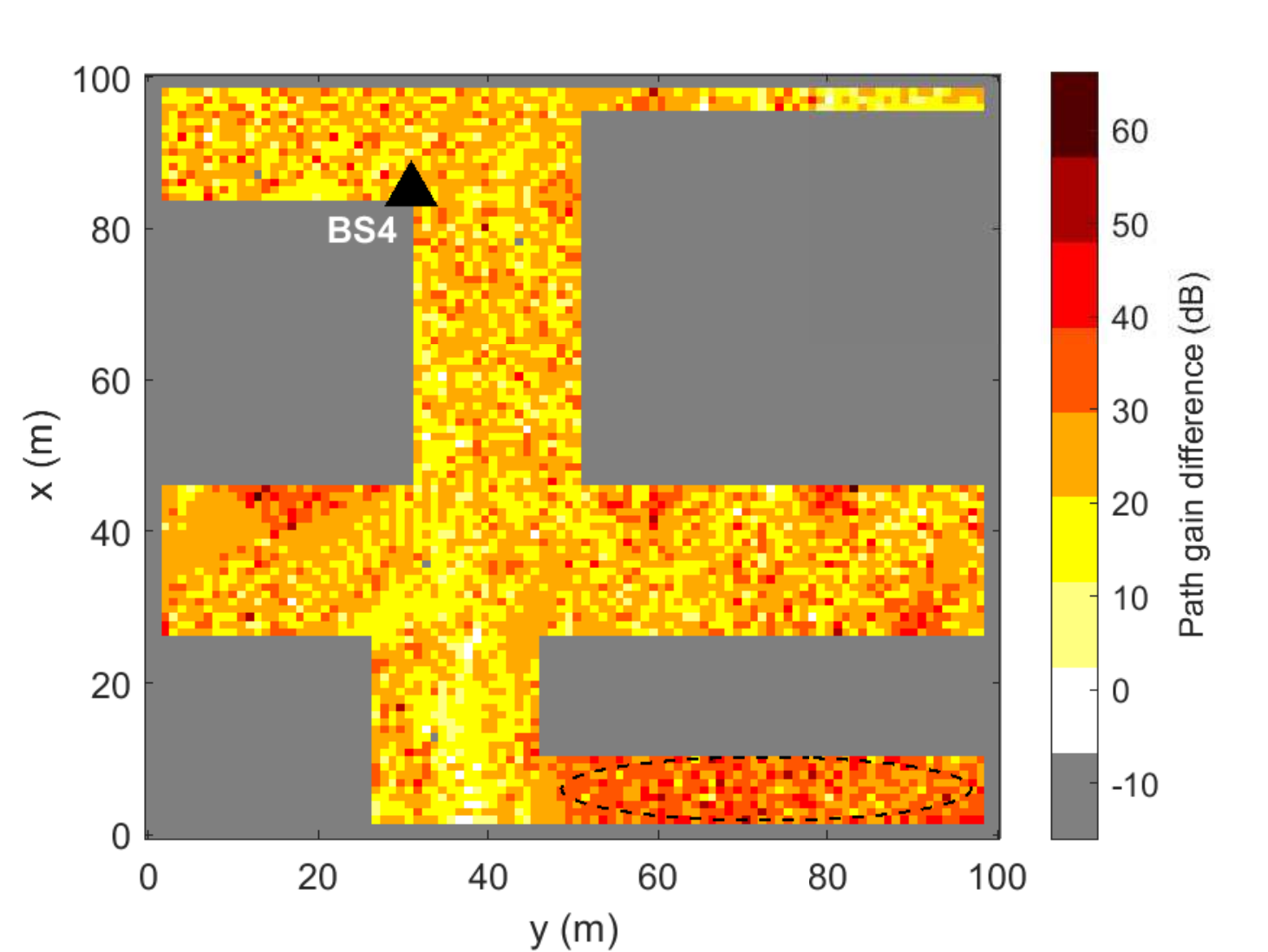}
\label{fig:PG_diff}}
\centering
\subfloat[]{\includegraphics[width=0.45\linewidth]{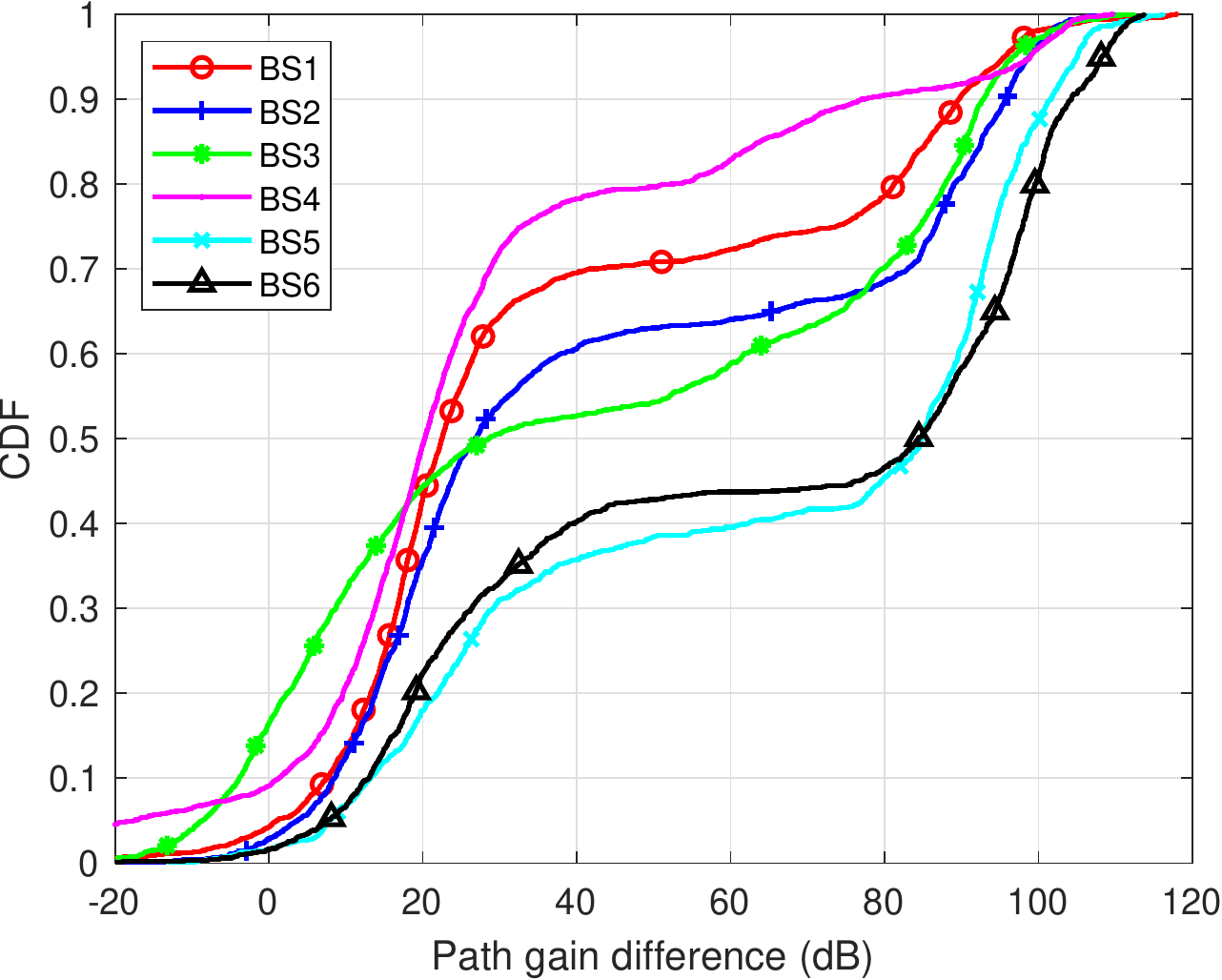}
\label{fig:CDF_PG_NLOS}}
\caption{(a) Heat map of the difference in path gain obtained at 2.4~GHz and 28~GHz when the BS is placed at position BS4. The warmer colors inside the dashed area represent higher difference in path gain due to blockages. (b) CDF of the difference in path gain at NLoS points for six BS locations.}
\label{fig:PGdiff}
\end{figure}

The rest of the article is organized as follows. First, the differences in the channel propagation characteristics of sub-6~GHz and mmWave bands are discussed. Following the definition of the mmWave BS placement problem, the MAB learning-based solution and action selection strategies are introduced. Finally, the results obtained with both the proposed method and the ray-tracing simulations are presented.

\section*{Propagation Channel Characteristics at Sub-6~GHz and mmWave Bands}
\label{sec:sub6-GHzVSmmWave}
A typical mmWave link suffers more than an order-of-magnitude larger path loss and is more susceptible to blockages than a traditional sub-6~GHz link~\cite{8895815}. These facts, in turn, raise several issues that should be addressed while planning the mmWave network. For example, to cope with the blockages, mmWave BSs are required to be densely deployed, which leads to interference issues and requires dedicated solutions, such as beamforming.

To illustrate the above-mentioned differences between sub-6~GHz and mmWave bands, we consider the urban outdoor environment of size 100~m $\times$ 100~m shown in Fig.~\ref{fig:process}. The building heights and widths are randomly selected in the range of (8~m, 25~m) and (20~m, 40~m), respectively. BSs are placed at six different locations, labeled as 1 to 6 at the bottom right in Fig.~\ref{fig:process}, on top edges of the buildings. Service area is defined as the whole area of interest minus the area covered by buildings. A receiver grid is placed on the service area with the antenna heights being equal to 1.5~m. Then, path gain at each receiver point on the service area is calculated using a ray-tracing software for a particular frequency from each band (i.e., 2.4~GHz and 28~GHz) by activating only one BS at a time and taking into account up to four reflections.\looseness=-1

Heat map of the path gain difference obtained at the two specified frequencies, when the BS is located at position BS4, is given in Fig.~\ref{fig:PGdiff}(a). The beamforming gains at mmWave frequencies are neglected for simplicity, which is easy to add for a given mmWave array size and beam pattern. Overall, the path gain is much higher at 2.4~GHz than at 28~GHz. An average difference of about 21~dB in the path gain is observed when the whole service area is considered. However, the difference is much larger at receiver locations where there is no line-of-sight~(LoS) to the BS and even larger, if there are also no low-order reflections (e.g., at locations inside the dashed ellipse). Fig.~\ref{fig:PGdiff}(b) shows the cumulative distribution function~(CDF) of the path gain difference when only the non-LoS~(NLoS) points are considered. In this case, the difference is more pronounced and as high as more than 80~dB on the average for locations BS5 and BS6. These observations point to the necessity of defining the mmWave BS placement problem by considering the typical properties of the mmWave systems.\looseness=-1

\section*{Problem Definition}
The problem of the optimal positioning of mmWave BSs can be defined as follows. We are given an outdoor environment and a set of possible BS locations. The goal is to find the best location(s) for the BS(s) that maximizes a QoS metric (e.g., coverage probability or throughput) in a predefined service area. The channel is time-varying due to several factors, such as mobile obstacles. Therefore, a potential solution should capture the stochastic channel behavior and identify the optimal BS location(s) that achieves the highest QoS in the long term.\looseness=-1

The DEM of a given environment is a 3D matrix with the first two dimensions being the x-y coordinates and the third dimension being the elevation. Each BS location in a finite set of possible locations is represented by its 3D spatial coordinates. The service area, where a predefined QoS metric(s) is/are desired to be achieved, is defined as the building-free area on the DEM and partitioned into grids. The number of grids is determined by the desired resolution, and each grid is represented by its center and the receiver height. 

For a given BS location, path loss at any grid can be estimated depending on the state of the link (i.e., LoS or NLoS) between the BS and the grid as~\cite{3GPP}:
\begin{equation}
\label{eq:3GPPLOS}
    L^{\mathrm{LoS}}=\:28+40\log_{10}(d_\mathrm{3D})+20\log_{10}(f_c)-9\log_{10}\left((d_\mathrm{BP})^2+(h_\mathrm{BS}-h_\mathrm{UT})^2\right)+X_{\sigma_{\mathrm{LoS}}},
\end{equation}
\begin{equation}
\label{eq:3GPPNLOS}
    L^{\mathrm{NLoS}}=\:13.54+39.08\log_{10}(d_\mathrm{3D})+20\log_{10}(f_c)-0.6(h_\mathrm{UT}-1.5)+X_{\sigma_{\mathrm{NLoS}}},
\end{equation}
where $d_\mathrm{3D}$ is the Euclidean distance in meters between the BS and the receiver or user terminal~(UT) in the corresponding grid, $f_c$ is the frequency in GHz, $h_\mathrm{BS}$ and $h_\mathrm{UT}$ are the heights of the BS and the UT, respectively, and $d_\mathrm{BP}$ in~(1) is the breakpoint distance that separates two different trends of path loss. The term $X_{\sigma}$ is a zero-mean lognormal random variable with standard deviation $\sigma_{\mathrm{LoS}}=4$ and $\sigma_{\mathrm{NLoS}}=6$ for LoS and NLoS links. This error term represents shadow fading and accounts for the time-varying channel behavior. Using the DEM of the environment, a viewshed analysis can be performed and incorporated into the path loss model, and as a result, path loss can be defined more generally as: 
\begin{equation}\label{eq:PL_combined}
    L=\delta\times L^{\mathrm{LoS}}+(1-\delta)\times L^{\mathrm{NLoS}},
\end{equation}
where the parameter $\delta$ represents the link state and takes the value 1 if the grid center is visible to the location of the associated BS and 0 otherwise.

The QoS metric is defined as the coverage probability, and only one BS is considered for simplicity. However, the MAB learning-based approach, which will be introduced in the next section, can easily be adapted to problems with multiple BSs and different QoS definitions. A grid is assumed to be in coverage if the path loss at that grid is below the maximum allowable path loss for reliable communication. Then, the probability of coverage within the intended area can be calculated as the ratio of the number of grids in coverage to the total number of grids.

\section*{MAB Learning-Based mmWave BS Placement}
This section gives a brief background on MAB learning and introduces the proposed mmWave BS placement method.
\subsection*{Background on MAB Learning}
MAB learning is an example of classic reinforcement learning~(RL), where there is a slot machine with a number of arms and each arm has its own probability of success. Each time the gambler pulls one of the arms, he/she gets an immediate stochastic \textit{reward} of either 1 (success) or 0 (failure). The objective is to identify which arm to pull to maximize the total reward collected in the long run. After its introduction in 1952~\cite{robbins1952some}, MAB learning has been used in various fields, such as tailoring content to Internet users based on their web history, or choosing treatments for patients based on their medical records~\cite{bastani2017}. MAB learning-based approaches have also attracted attention in recent years in wireless communications, for example in beam selection, beam alignment and tracking, and throughput enhancement~\cite{booth2019}.

The non-triviality of the MAB problem lies in the fact that the probability distribution of the reward corresponding to each arm is unknown to the gambler. Therefore, the learning can only be carried out via means of trial-and-error. Given these, the MAB problem can be considered as a single-step Markov decision process with each arm selection corresponds to an \textit{action}, which is then followed by an immediate reward. Similar to the classical MAB problem, since the state of the link between a point in the service area (or a user) and the associated BS, the blockages, and many other channel parameters are all stochastic, the rewards (QoS levels) are also stochastic in the BS placement problem. 
\subsection*{Proposed Method}
The following analogy is used between the mmWave BS placement problem and the classical MAB problem. Possible BS locations correspond to the arms of the slot machine. The action of choosing a particular BS location among possible locations corresponds to pulling one of the arms of the slot machine. The immediate reward of choosing a BS location is the stochastic QoS metric returned (i.e., the coverage probability). Finally, the \textit{action-value} (or the Q-value) of a BS location is the expected reward and can be estimated by averaging the coverage probability obtained for that location up to the current time step. The goal is to find the best \textit{policy} or the optimal BS location that maximizes the cumulative reward at the end of the training process. \looseness = -1

One possible approach to the MAB problem is using the classical RL technique of an $\epsilon$-greedy \textit{agent}. According to the $\epsilon$-greedy action selection strategy, at each iteration, the agent chooses a random BS location with a small probability and the best BS location learned so far otherwise. Following the BS selection, path loss at each point in the service area is calculated using either the LoS or NLoS model based on the viewshed analysis. The viewshed analysis determines the visibility of a point to a BS location by finding the shortest path linking two nodes in the BOMs and checking if the path contains any binary ones~\cite{BOM_2015}. By comparing the path loss with a predefined threshold, coverage probability in the service area and the Q-value corresponding to the current BS location choice is calculated.\looseness=-1

\begin{figure}[t]
\vspace{-2mm}
\centering
\subfloat[]{\includegraphics[width=0.45\linewidth]{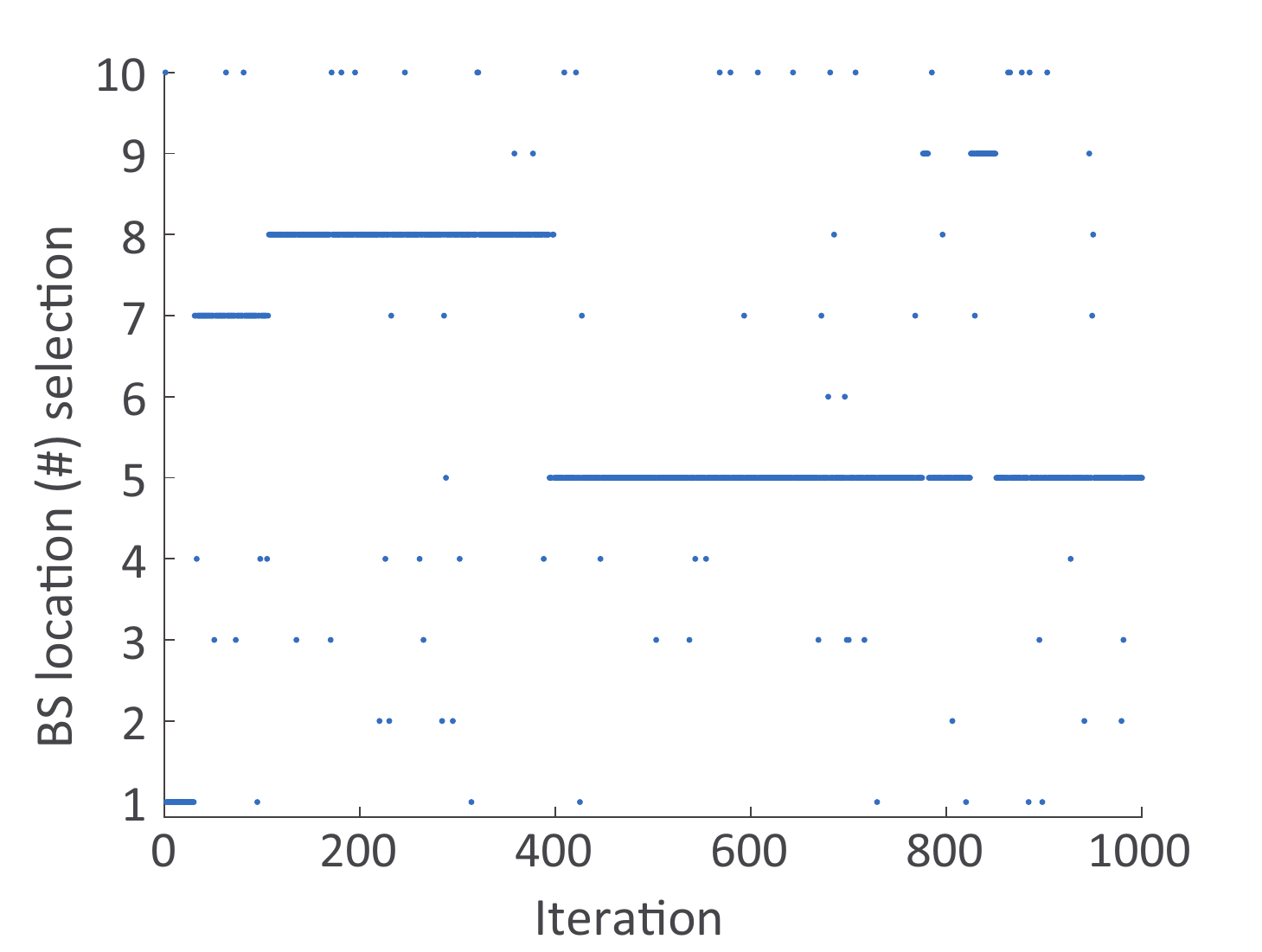}
\label{fig:BSSelection_EpsGreedy}}
\subfloat[]{\includegraphics[width=0.45\linewidth]{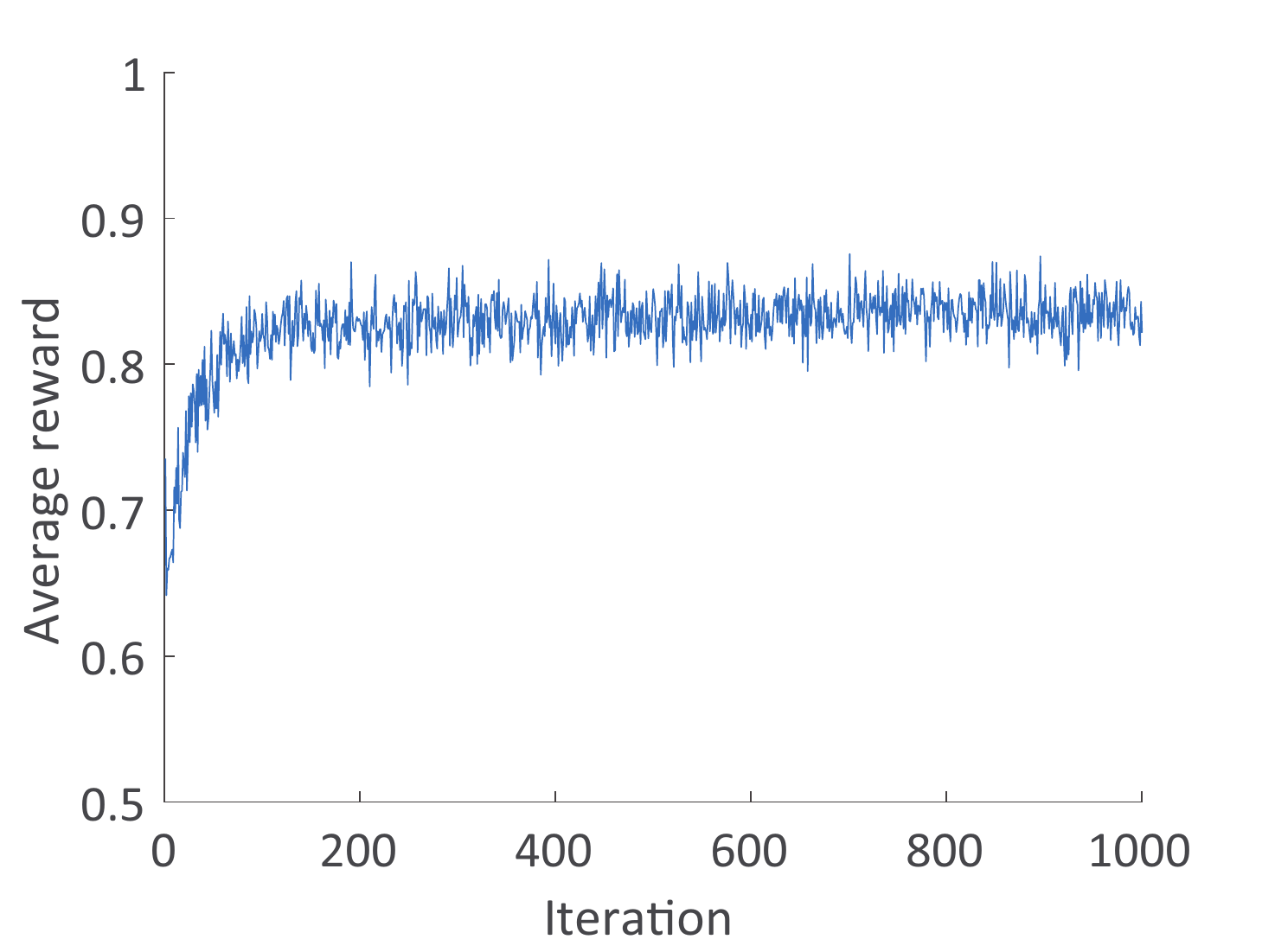}
\label{fig:AveragedRewards}}
\caption{(a) BS location selection for a single episode using the $\epsilon$-greedy strategy. (b) Reward vs. iterations (averaged over 50 episodes).}
\label{fig:BSChoice_AvgRew}
\end{figure}

Each time the agent picks the same BS location, the Q-value of that location is updated based on the coverage probability obtained. As time passes, the agent gets to better know the coverage performance of each BS location under the time-varying channel and takes better actions to collect more rewards. BS location selections (among the 10 BS locations shown in Fig.~\ref{fig:process}) for a sample episode is given in Fig.~\ref{fig:BSChoice_AvgRew}(a). In the first few hundred iterations, the agent chooses the suboptimal locations (i.e., BS7 and BS8) more frequently. However, after having some interaction with the environment and experiencing the consequences of its choices, it learns the optimal location BS5 and sticks to it around after 400 iterations. Fig.~\ref{fig:BSChoice_AvgRew}(b) shows the average reward collected through iterations. The average reward is around 0.6 during the first iterations. After about 100 iterations, the algorithm learns the optimal location and achieves an average reward of about 0.83 by choosing the optimal location more often.\looseness=-1

Fig.~\ref{fig:BSChoice_AvgRew}(a) shows the BS selections for a single episode where the agent successfully finds the optimal BS location. However, in some other episodes, due to the time-varying channel, sometimes it may turn out that the path gain on the service area is lower when the BS is placed at the global optimal location than when it is placed at a suboptimal location. If this situation repeats for a number of iterations, the algorithm may get stuck in a suboptimal location as the Q-value corresponding to that location will be slightly larger than that of the global optimum. This problem can easily be handled by increasing the amount of training data or the number of episodes.\looseness=-1

To find the best policy, the agent needs to gather enough information for the best overall decisions, and, on the other hand, it needs to keep the risk associated with the bad actions at a reasonably low level. This is known as the \textit{exploitation-exploration} dilemma. With exploitation, the agent acts greedily and takes advantage of the best option it knows so far. With exploration, the agent takes short-term risks to collect information regarding the consequences of taking unknown actions. The next section describes different exploration strategies to solve the defined MAB problem. 

\vspace{-2.5mm}

\subsection*{Exploration Strategies}
Choosing the way that the agent explores the actions and their stochastic outcomes through the iterations is critical for a given MAB problem. In the \textit{$\epsilon$-greedy} approach, it is the value of $\epsilon$ that determines whether a random or the current best location to be selected at each time step. Instead of setting the $\epsilon$ value to a constant, it may be kept dependent on time. This strategy is known as the \textit{decayed $\epsilon$-greedy}. With this strategy, the value of $\epsilon$ is reduced as time passes so that the agent relies more on its experiences and takes random actions less often. \looseness=-1 

Although both the $\epsilon$-greedy and decayed $\epsilon$-greedy algorithms provide a balance between exploration and exploitation, while exploring the environment, the worst action and any other action are equally likely to be chosen. To address this issue, the \textit{softmax exploration} strategy increases the frequency of choosing the actions that might return higher rewards by varying the action probabilities based on the estimated Q-values. The most common softmax method uses a Boltzmann distribution~\cite{RL_book}. 

While exploiting the knowledge, all the action selection strategies discussed so far select the action with the highest Q-value even when there is only a slight difference between the Q-values of the first two best ranked actions. \textit{Upper confidence bound~(UCB)}-based approaches handle this problem by defining an upper bound for each action and exploring all the actions until being confident about each. For example, the \textit{UCB1} method chooses the action that yields the maximum value for the sum of the Q-value and a \textit{confidence} term~\cite{Auer_UCB1}. The confidence term is defined as a function of number of times a particular action is selected so far so that the uncertainty about that action decreases with the number of times it is chosen. With this approach, all possible BS locations are eventually selected and explored sufficiently. As time passes, Q-value dominates the confidence term, and the UCB1 method repeatedly selects the optimal BS location. 

Before presenting the experimental results, we note that the BS placement problem defined above can also be solved offline using brute-force search with the expense of exponential computational complexity, especially when the number of BSs and/or the size of the environment increases. However, exhaustive search is not a solution in real-time applications, such as sleep mode optimization. Spatial and temporal changes in daily traffic demand occur due to diverse applications, types of user equipment, and change in user densities. For the sake of energy saving, underutilized BSs may be switched off at low-traffic periods (e.g., night time). Although the BS placement problem can be solved by both MAB learning and brute-force search, dynamic on/off decision based on varying traffic demand requires fast converging algorithms for which the proposed MAB learning framework is a proper choice and the action selection strategies are critical.

\section*{Experimental Results }
\label{sec:Results}
In this section, we present the results of MAB learning-based mmWave BS placement method and compare them with those of the ray-tracing simulations. The BS antennas are assumed to use the 28~GHz frequency band. The maximum allowable path loss is set to 110~dB. 

For the outdoor urban environment in Fig.~\ref{fig:process}, first, the DEM is created. Possible BS locations are defined as the points on roof-top edges of the buildings in the area of interest (see the points marked by red asterisk in the DEM). Then, BOM is generated from the DEM with a grid resolution of 1~m $\times$ 1~m. The scenario, without loss of generality, is simplified to a single BS. Next, for all possible BS locations, the viewshed analysis is performed and the points in the service area that are visible to each BS location are identified. Assuming any two neighboring BS locations will likely have similar coverage performance, considering all the possible locations would be time consuming. Besides, it is less likely that the optimal BS location will be on the edges that are too close to the boundaries of the environment. Therefore, we discard those edges and only pick the location that has the maximum viewshed area from each of the remaining edges. The resulting 10 candidate BS locations are shown at the bottom right in Fig.~\ref{fig:process}. The candidate BS locations can be identified using other methods in the literature, for example, the computational geometry-based method in~\cite{Yanikomeroglu_2017}.
\begin{figure}[t]
\centering
\centerline{\includegraphics[width=0.6\linewidth]{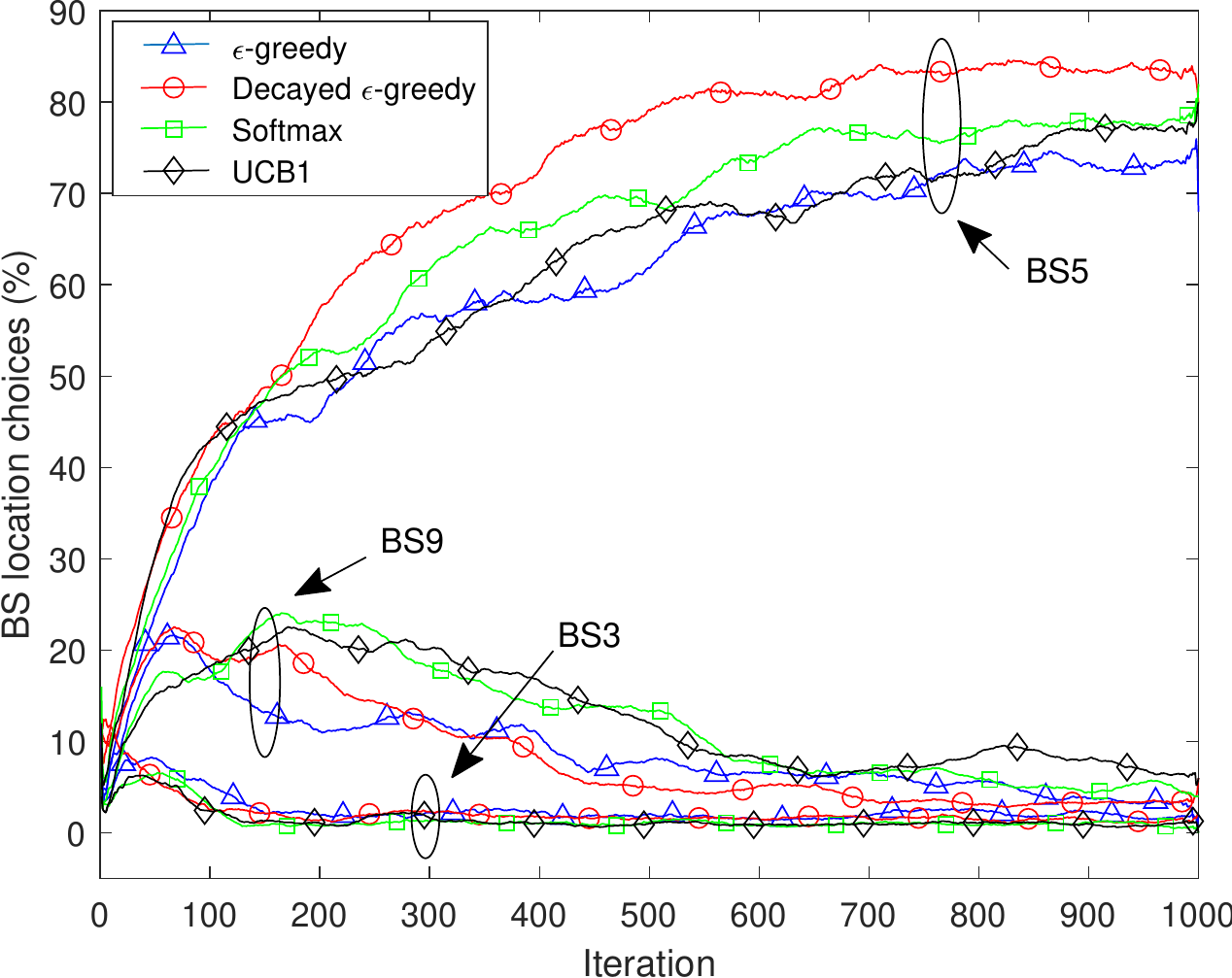}}
\caption{Number of times BS3, BS5, and BS9 are selected in percentages through iterations for different action selection strategies. Results are averaged over 50 episodes.}
\label{fig:BSchoices_inPercent}
\end{figure}


\subsection*{Comparison of Action Selection Strategies}
To find the optimal BS location, we use MAB learning with different action selection strategies described in the previous section. To average out the bias introduced by the outlier channel parameters, 50 episodes consisting of 1000 iterations are considered for each strategy. BS selection percentages through iterations are presented in Fig.~\ref{fig:BSchoices_inPercent} for three candidate BS locations, namely, BS3, BS5, and BS9. For all strategies, the optimal BS location is found to be BS5 with a selection frequency of about 80 percent after convergence, and it is followed by the location BS9. On the other hand, it is agreed that the location BS3 is not a good choice to place the BS, and it is almost never selected as the learning progresses. 

Decayed $\epsilon$-greedy method finds the optimal location faster than the $\epsilon$-greedy method. This is because, in the case of the former, the value of $\epsilon$ drops below a small probability value after a few hundred iterations, and after that, the actions with smaller Q-values are selected less often. The convergence rates are similar in UCB1 and softmax exploration methods. Moreover, the relative frequency of being selected is higher for the suboptimal locations when UCB1 method is used. This verifies the motivation of the UCB1 method, that is, it keeps exploring the locations until the uncertainty about each drops below a certain level, and then it selects the candidate locations that yield higher Q-values more frequently. 

It is pertinent to note that we only considered the shadow fading term in the path loss equations to represent the stochastic characteristics of the channel. However, it is straightforward to consider other random parameters, such as nonuniform and/or time-varying user distribution. In such a case, MAB learning will take those parameters into account while calculating the rewards and decide the optimal BS location accordingly. In addition, the MAB learning method only aims to maximize the given QoS metric, which is the coverage probability in this article. However, the maximum QoS achieved by placing the BS at the optimal location may not meet the expectations. In such a case, number of BSs can be increased until the desired QoS level is reached. \looseness=-1

\begin{figure*}[t]
\centering
\centerline{\includegraphics[ width=0.65\linewidth]{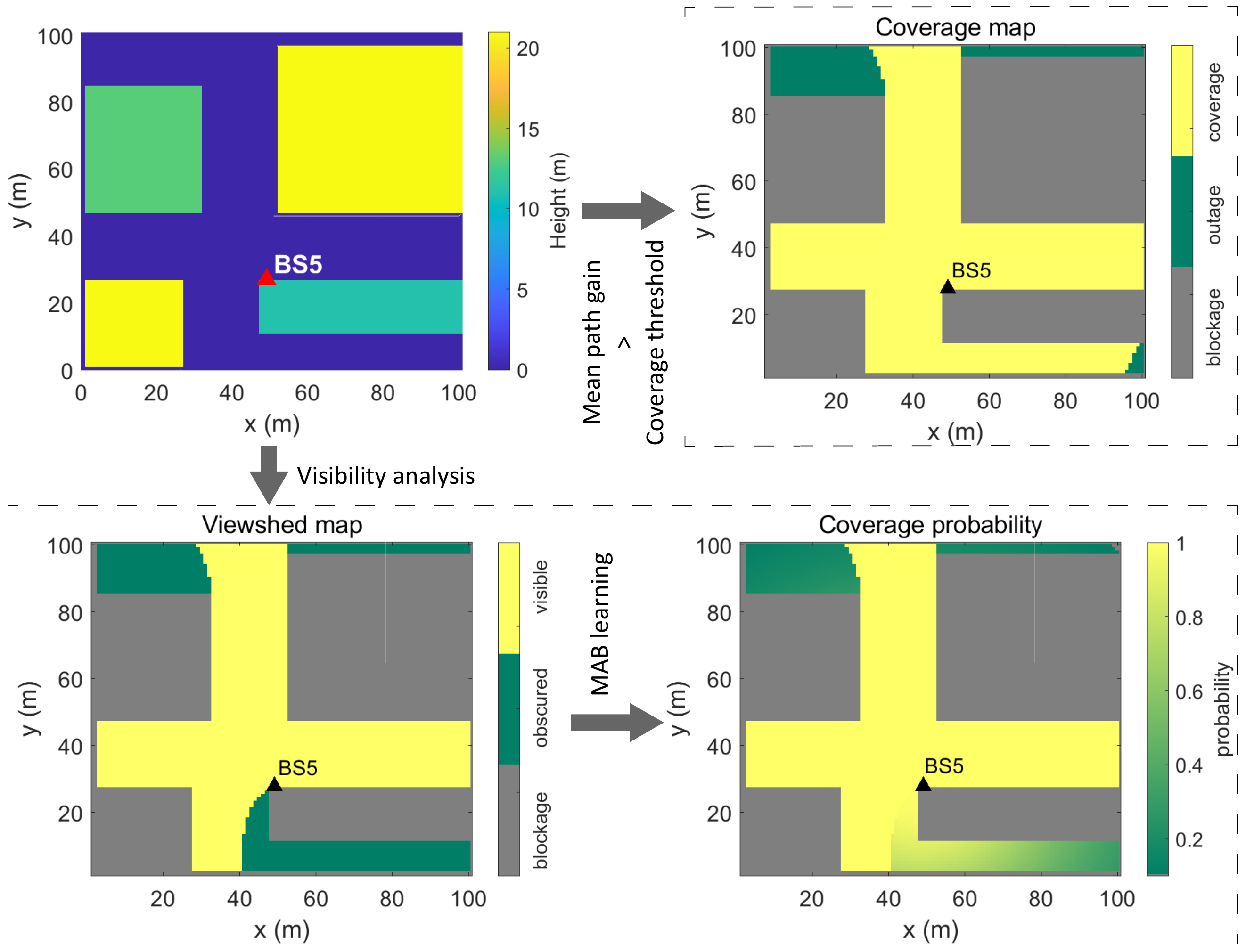}}
\caption{Coverage estimation in the service area based on the mean channel statistics and the MAB learning approach.}
\label{fig:viewshed_VS_coverage}
\end{figure*}

\subsection*{Mean Channel Statistics vs. MAB Learning}
Next, we compare the coverage in the service area estimated using mean channel statistics and the proposed MAB learning-based approach. The results, when the BS is placed at the optimal position BS5, are presented in Fig.~\ref{fig:viewshed_VS_coverage}. The coverage map at the top right of the figure is constructed based on the mean path gain at the grid centers and comparing it to the coverage threshold. The mean gain at a grid is calculated by averaging the linear path gains obtained from the iterations when the optimal BS location is selected. On the other hand, the coverage probability (shown below the coverage map) is calculated for each grid by averaging the binary coverage results obtained from the iterations when the optimal BS location is selected. The shaded region represents the grids that have a coverage probability different than zero or one. That is, due to the temporal changes in the channel, those grids are sometimes in coverage and sometimes in outage. This result cannot be observed in the coverage map. Therefore, if one were to rely on the averaged path gain to determine the optimal BS location, then it would not be possible to capture the instantaneous variations in the channel, and the BS location determined might not be guaranteed to be optimal in terms of long-term QoS.

In Fig.~\ref{fig:viewshed_VS_coverage}, it is also observed that, as expected, there is a high correlation between the visibility and the coverage probability. The viewshed area is a subset of the coverage area because the viewshed analysis cannot consider the contribution of reflections from buildings to the path gain. However, the above-mentioned relation between the visibility and the coverage 
may not always hold. For example, for an area that is larger than the one considered here, the path loss at larger distances can be above the allowable threshold even if the grids at those distances lie in the viewshed area. Therefore, having a greater viewshed area for a BS location does not necessarily imply that the coverage performance will be better for that BS location when compared to the other candidate locations. \looseness=-1


\subsection*{Comparison with Ray-Tracing Simulations}
In this section, we compare the coverage performance of each candidate BS location estimated with the MAB learning-based approach and the ray-tracing simulations. We note that, in ray-tracing simulations, it is not possible to account for the dynamic components of the environment which are already shown to affect the channel behavior and hence the decision of the optimal BS location. In addition, since the path loss models used in MAB learning are estimated from a limited number of measurements, path loss obtained from the models and the ray-tracing will differ from each other. However, it is still reasonable to expect some consistencies between the results of the two approaches at least. \looseness = -1

To make a fair comparison between the two approaches, while creating the environment in the ray-tracing software, we added the buildings that surround the area of interest as shown in Fig.~\ref{fig:process}. This way, the reflections from those buildings are taken into account while calculating the path gain, as in the case where 3GPP path loss models are used. In ray-tracing simulations, half-wave dipole antennas are used both at the BS and the receiver points. Ray spacing is set to 0.50$^\circ$, and the maximum number of reflections is limited to four.


\begin{figure}[t]
\centering
\centerline{\includegraphics[trim=1.5cm 0cm 0cm 0.45cm, clip, width=0.6\linewidth]{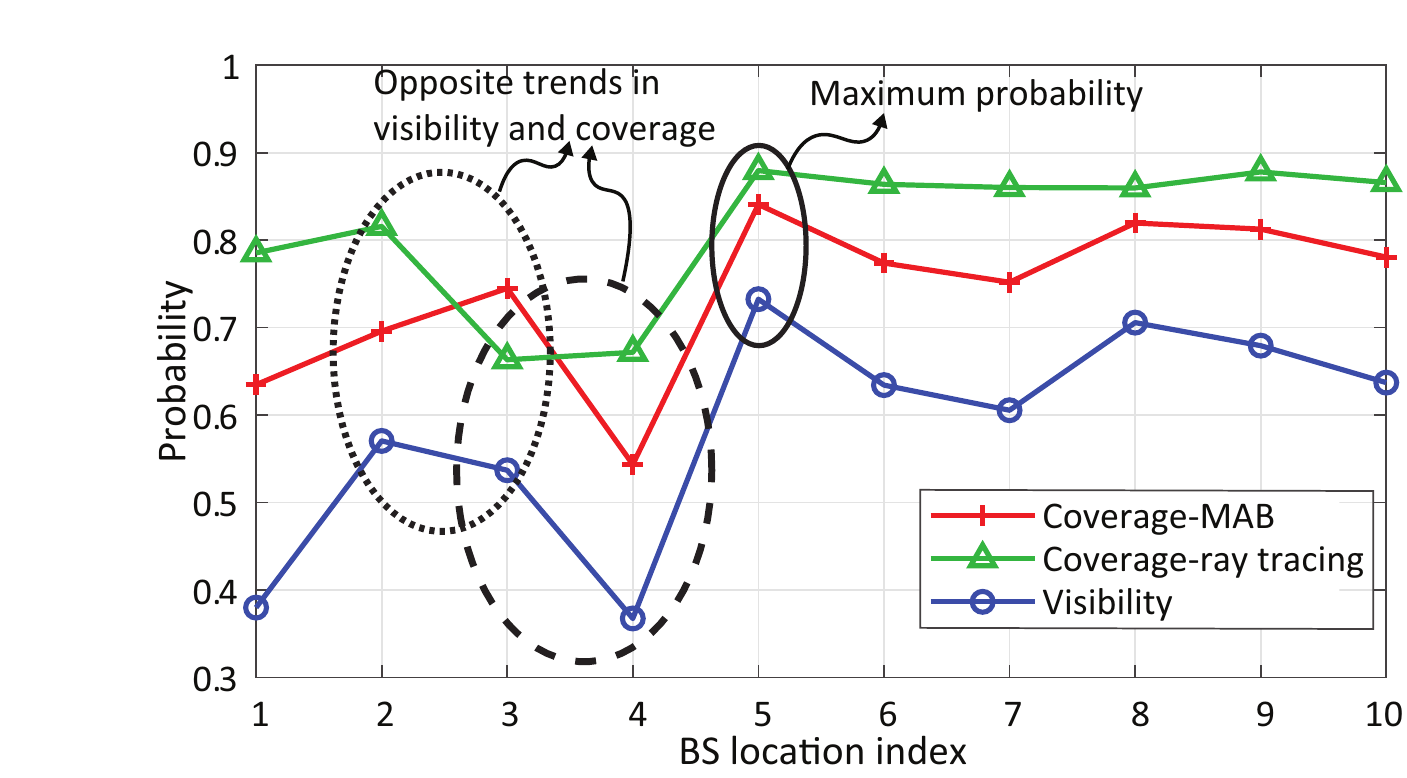}}
\caption{Coverage probability obtained from the (averaged Q-values of the) MAB learning and the ray-tracing simulations, and the visibility probability. BS location index refers to the locations in Fig.~\ref{fig:process}.}
\label{fig:MABvsRayTracingvsVisib}
\end{figure}

In the ray-tracing simulator, we activate only one BS at a time to compute the coverage rate for the whole service area. On the other hand, for the MAB learning, we calculate the coverage probability based on the averaged Q-value of each candidate BS location at the end of the training process. The coverage probability along with the visibility probability for each candidate BS location are presented in Fig.~\ref{fig:MABvsRayTracingvsVisib}. Overall, we observe a strong correlation between the coverage and the visibility. The ray-tracing simulations and the MAB learning demonstrate the same trend in the results (except for the locations BS3 and BS4), and both return the location BS5 as the optimal BS location and the locations BS8 and BS9 as the runner-ups. These results are in accordance with the results provided in Fig.~\ref{fig:BSchoices_inPercent} where the relative frequency of being selected are shown to be higher for the locations BS8 and BS9. \looseness=-1

The dotted and the dashed circles in Fig.~\ref{fig:MABvsRayTracingvsVisib} label the BS locations where the trends in the coverage and the visibility are in opposite directions. Although the visibility decreases when moving from the location BS2 to the location BS3, the expected coverage estimated by the MAB-learning increases. Similarly, the coverage probability obtained from the ray-tracing slightly increases when moving from BS3 to BS4 as opposed to steep decrease in the visibility. These results justify the use of the MAB learning-based approach as a more reliable method in planning the mmWave network than the methods that rely only on the viewshed analysis or the methods that use the mean statistics of the channel in an optimization framework.

\section*{Conclusion}
\label{sec:Conclusion}
In this article, we introduced a novel method for mmWave BS placement using MAB learning. The proposed method incorporates viewshed analysis and unique characteristics of mmWave bands while determining the coverage in the service area. By relying on the 3D models of the environment and taking the temporal changes in the channel into consideration, it is shown to be more effective for maximizing the long-term coverage performance than the geometry and optimization-based methods. Representative results are presented for the placement of a single BS case. However, the approach used can easily be adapted such that it finds the optimal configuration in the case of multiple BSs and for real-time applications, such as sleep mode optimization of mmWave BSs based on time-varying traffic conditions. Similarly, coverage performance can be replaced by any other QoS definition without changing the current framework. In addition, by creating appropriate BOMs, the proposed method can also be used in indoor environments. 

\section*{Acknowledgment}
This work has been supported in part by NSF through CNS-1916766.
\bibliography{IEEEabrv,references}
\bibliographystyle{IEEEtran}

\end{document}